\documentstyle[aps]{revtex}
\begin{document}

\draft
\twocolumn
\title{Spin and charge gaps in the one-dimensional Kondo-lattice 
model with \\  
Coulomb interaction between conduction electrons}

\author{Naokazu Shibata} 
\address{
Institute for Solid State Physics, University of Tokyo,  
7-22-1 Roppongi, Minato-ku, Tokyo 106, Japan, \\
and Department of Physics, Faculty of Science,
Science University of Tokyo, 
1-3 Kagurazaka, Shinjuku-ku, Tokyo 162, Japan 
}
\author{Tomotoshi Nishino}
\address{
Department of Physics, Graduate School of Science, 
Tohoku University, Sendai 980, Japan
}
\author{Kazuo Ueda}
\address{
Institute for Solid State Physics, University of Tokyo, 
7-22-1 Roppongi, Minato-ku, Tokyo 106, Japan
}
\author{Chikara Ishii}
\address{
Department of Physics, Faculty of Science, 
Science University of Tokyo, 
1-3 Kagurazaka, Shinjuku-ku, Tokyo 162, Japan
}

\date{26 Dec 1995}

\maketitle

\begin{abstract}
The density-matrix renormalization-group
method is applied to the one-dimensional Kondo-lattice model
with the Coulomb interaction between the conduction electrons.
The spin and charge gaps are calculated as a function of 
the exchange constant $J$ and the Coulomb interaction $U_c$.
It is shown that both the spin and charge gaps increase with 
increasing $J$ and $U_c$. 
The spin gap vanishes in the limit of $J \rightarrow 0$
for any $U_c$ with an exponential form,
$\Delta_s\propto \exp{[-1/\alpha (U_c) J \rho]}$.
The exponent, $\alpha (U_c)$, is determined as a function of $U_c$. 
The charge gap is generally much larger than the spin gap.
In the limit of $J \rightarrow 0$, the charge gap vanishes as
$\Delta_c=\frac{1}{2}J$ for $U_c=0$ but for a finite $U_c$ 
it tends to a finite value, which is the charge gap
of the Hubbard model.
\end{abstract}


Recently the insulating phase of the Kondo-lattice model (KLM) has 
been extensively studied in relation to the Kondo insulators 
\cite{aeppli}.
In contrast to the strong-coupling limit where various 
properties are easily understood from the local bases,
the weak-coupling limit has not yet been fully understood.
Since the KLM with weak exchange coupling is an effective model 
of the symmetric periodic Anderson model with strong
Coulomb interaction, the weak-coupling region of the KLM 
is important to understanding the strongly correlated insulating 
phase. For the one-dimensional case it has been established that 
both spin and charge gaps exist for any exchange coupling.
The excitation gaps vanish in the limit of vanishing exchange 
coupling \cite{tsunetsugu,tsvelik}. 
Since the gaps are tiny in the weak-coupling regime,
it is no longer justified to neglect Coulomb interaction 
between the conduction electrons. 

In this article we determine the spin and 
charge gaps in the entire region of exchange constant
taking account of the Coulomb interaction
between the conduction electrons $U_c$. 
The excitation gaps are obtained precisely by employing the
density matrix renormalization group (DMRG) method.
Among other things, we find that $U_c$ actually stabilizes the 
Kondo spin liquid phase, resulting in monotonic increase of 
both the spin and charge gaps as a function of $U_c$.

The model we consider in this article is the following
one-dimensional KLM with Coulomb interaction (KLMC):
\begin{eqnarray}
H & = & -t\sum_{i \sigma}
( c_{i \sigma}^\dagger c_{i+1 \sigma} + \mbox{H.c.})  
+J \sum_{i \mu} S^\mu_i \sigma^\mu_i \nonumber\\
 &  & + \ U_c\sum_i (c_{i \uparrow}^\dagger 
c_{i \uparrow}-\frac{1}{2}) (c_{i \downarrow}^\dagger 
c_{i \downarrow} -\frac{1}{2}) ,   
\end{eqnarray}
where $c_{ i \sigma}^\dagger \  (c_{ i \sigma}) $ is the creation 
(annihilation) operator of a conduction electron  
at the $i$th site, and
$ \sigma^\mu_i=(1/2)\sum_{\sigma\sigma'}
c_{i\sigma}^{\dagger} \tau^\mu_{\sigma\sigma'}c_{i\sigma'} $
with the Pauli matrices $\tau^\mu_{\sigma\sigma'}$ ($\mu = x,y,z$)
are the spin
density operators of the conduction electrons.
In the second term $ S^\mu_i $ represents a localized 
$f$-spin operator with $S=1/2$.  
Concerning the kinetic energy of the conduction 
electrons we consider only the hopping processes between the 
nearest-neighbor sites, $t$.  
This model was considered by Yanagisawa and Harigaya
in connection with the ferromagnetic
ground state in the strong-coupling limit of the one-dimensional
KLM away from half-filling \cite{yanagisawa}.
Here we are interested in the half filling case where the total 
number of conduction electrons is equal to the number of lattice 
site: $N\equiv \sum_{i \sigma} 
c^{\dagger}_{i \sigma}c_{i \sigma}=L$.
It is needless to say that this Hamiltonian is reduced to 
the Hubbard model in the limit of vanishing exchange interaction, 
$J\rightarrow 0$, and to the usual KLM for $U_c=0$.

For the case of $U_c=0$ extensive studies 
on the spin and charge gaps have already been done.
First Tsunetsugu $et\ al.$ showed numerically that 
the ground state is singlet and the spin gap always 
exists for any finite exchange by using exact 
diagonalization combined with finite-size scaling 
\cite{tsunetsugu}.  
They have concluded that the $J$ dependence of the spin 
gap has a similar
form as the Kondo temperature of the single 
impurity Kondo model but the exponent is 
larger than that of single impurity case, 
the lattice enhancement effect.
Subsequently, semiclassical analysis based on the 
nonlinear $\sigma$ model obtained by mapping from
the one-dimensional
KLM was carried out by Tsvelik \cite{tsvelik}.
He showed that the spin gap 
exists for any $J$ in consistent with the above results with some 
logarithmic correction in the exponent.
For the charge gap, on the other hand, 
a $J$-linear behavior is suggested by Nishino and Ueda.
By using exact diagonalization of the one-dimensional 
periodic Anderson model up to eight sites, they observed that 
the charge gap is proportional to $1/U$ 
in the region of strong Coulomb interaction 
which means $J$-linear dependence in the corresponding
weak-coupling region of the KLM ($J=8V^2/U$) \cite{nishino}.
It follows that the ratio between the 
charge and spin gaps $\Delta_c / \Delta_s$ 
diverges in the limit of vanishing exchange constant.
Yu and White applied the DMRG method to the KLM \cite{yu}.
Their results for larger clusters have confirmed the diverging 
behavior of $\Delta_c / \Delta_s$ 
in the region of weak exchange coupling.
For an asymmetric case, however it is also pointed out the 
ratio remains finite \cite{guerrero}.

In spite of these progresses, the results in the region of 
weak exchange coupling are still not accurate enough to 
determine precisely the functional form of the gaps due to the 
smallness of clusters used in these numerical studies.
For example, the exponent of the spin gap
has not yet been determined with sufficient accuracy
and the linear $J$ dependence of the charge gap
remains to be a plausible scenario.
In order to determine functional forms of these gaps
it is essential to treat 
large clusters systematically so that we can
extrapolate the gap energies to the bulk limit.
For this purpose the DMRG method developed by White \cite{white}
is efficient because the number of the 
states used to construct the wave function
does not increase with increasing the system size.
Naturally the truncation of the states introduces 
numerical errors, but the errors may be estimated by the 
eigenvalues of the density matrix which are truncated off.
Thus it is possible to increase the system size
within a given accuracy.

In the previous DMRG calculation by Yu and White \cite{yu}, 
the number of states kept for each block is 180 and 
the number of the maximum system size is 24.
However the lattice sizes are too small to fix the 
$J$ dependence of the spin and charge gaps.
For this purpose 
we use the finite system algorithm 
of the DMRG method with open boundary conditions 
keeping up to 300 states for each block and increase 
the system size up to 80.

In contrast to the infinite system algorithm, the
finite system algorithm gives more accurate results,
but the extrapolation to the infinite system is necessary.
The extrapolation to the infinite may be done
in the following way.
Since the lowest excited state generally corresponds to the
bottom of an excitation spectrum which can be expanded in 
terms of $k^2$, 
we expect finite-size scaling is of the form of $L^{-2}$ 
for large systems.
It means the gap energy of the large systems behaves as
\begin{equation}
\Delta(L)=\Delta(\infty)+\beta L^{-2}+O(L^{-4}) .
\label{finite}
\end{equation}
We determine the gap 
energy of the bulk system by using this scaling 
when data for available system 
sizes already follows the scaling form.
When a gap is tiny, for example, the spin gap of $J=0.6t$ 
and $U_c=0$, the scaling form is not yet clear for the
system size available. In such a case 
the upper and lower bounds are estimated by 
$L^{-2}$ and $L^{-1}$ scaling, respectively, because 
the size dependence of the gap for smaller systems is 
close to $L^{-1}$ rather than $L^{-2}$.

First we discuss the spin gap.
It is obtained from the 
difference of the ground-state energies in the subspace of
total $S^z$ being zero and one
with the same total electron number $L$,
$E_g(S^z=1,N=L)-E_g(S^z=0,N=L)$. 
The SU(2) symmetry in the spin space guarantees the energy
difference  is the same as the spin gap in the subspace of total 
$S^z$ being zero.

Before proceeding to the analysis of the results
it is worth reminding the Kondo impurity model.
In the Kondo impurity model all physical quantities are scaled by 
single energy scale $T_K \sim D\exp{(-\frac{1}{\rho J})}$  
which is known as the Kondo temperature. 
Here $\rho = \frac{1}{2\pi t}$ is the density of states 
of the conduction band at the Fermi level.
The simplest extension to the lattice problem is to include an
enhancement of the exponent owing to intersite correlations.
Then the spin gap is expected to behave as 
\begin{equation}
\Delta_s \propto \exp{(-\frac{1}{\alpha \rho J})},
\label{s_gap}
\end{equation}
where the $\alpha$ is the enhancement factor.
The Gutzwiller approximation predicts an enhancement factor of 
2 \cite{rice}.
As a matter of fact, Tsunetsugu $et\ al.$\  have\ estimated 
that the enhancement factor $\alpha$ is in the range of  
$1\le \alpha \le 5/4$ by using a finite-size scaling based on 
exact diagonalizations for the systems up to $L=10 $. 

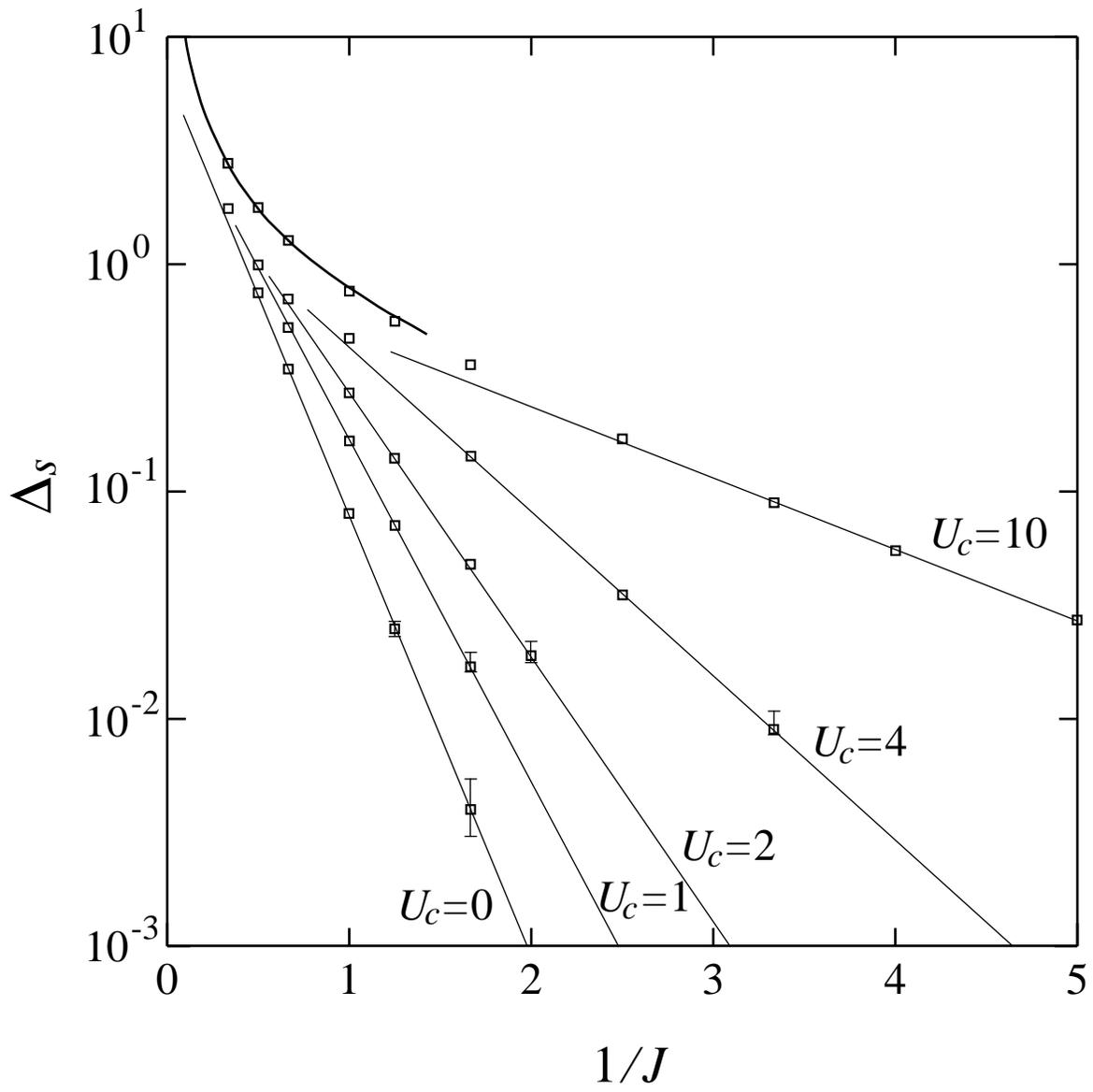
\begin{figure}
\caption{Spin gap of the one-dimensional Kondo lattice model
with Coulomb interaction. The thick curve represents the 
result of the perturbation theory in terms of $t/J$ for $U_c=10t$.
Typical truncation error in the DMRG calculation is 
$10^{-6}$  for $J=1$.
Errorbars are estimated from $L^{-1}$ and $L^{-2}$ scaling. 
Gap energies, exchange constant $J$, and 
Coulomb interaction $U_c$ are in units of $t$.
}
\label{spin_gap}
\end{figure}

The first task of the present study is to determine 
this enhancement factor more precisely.
For this purpose we plot logarithm of the spin gaps as a 
function of $1/J$. Figure \ref{spin_gap} shows the results 
extrapolated to the bulk limit using data of $L=6,8,12,18,24,40$.
From this figure we obtain the exponent $\alpha =1.4(1)$
for the case of $U_c=0$.
There are some ambiguities in the extrapolation 
to the bulk limit for a tiny gap. However within the present 
accuracy we do not observe any logarithmic correction to the 
exponent which is predicted by the semiclassical 
approach.

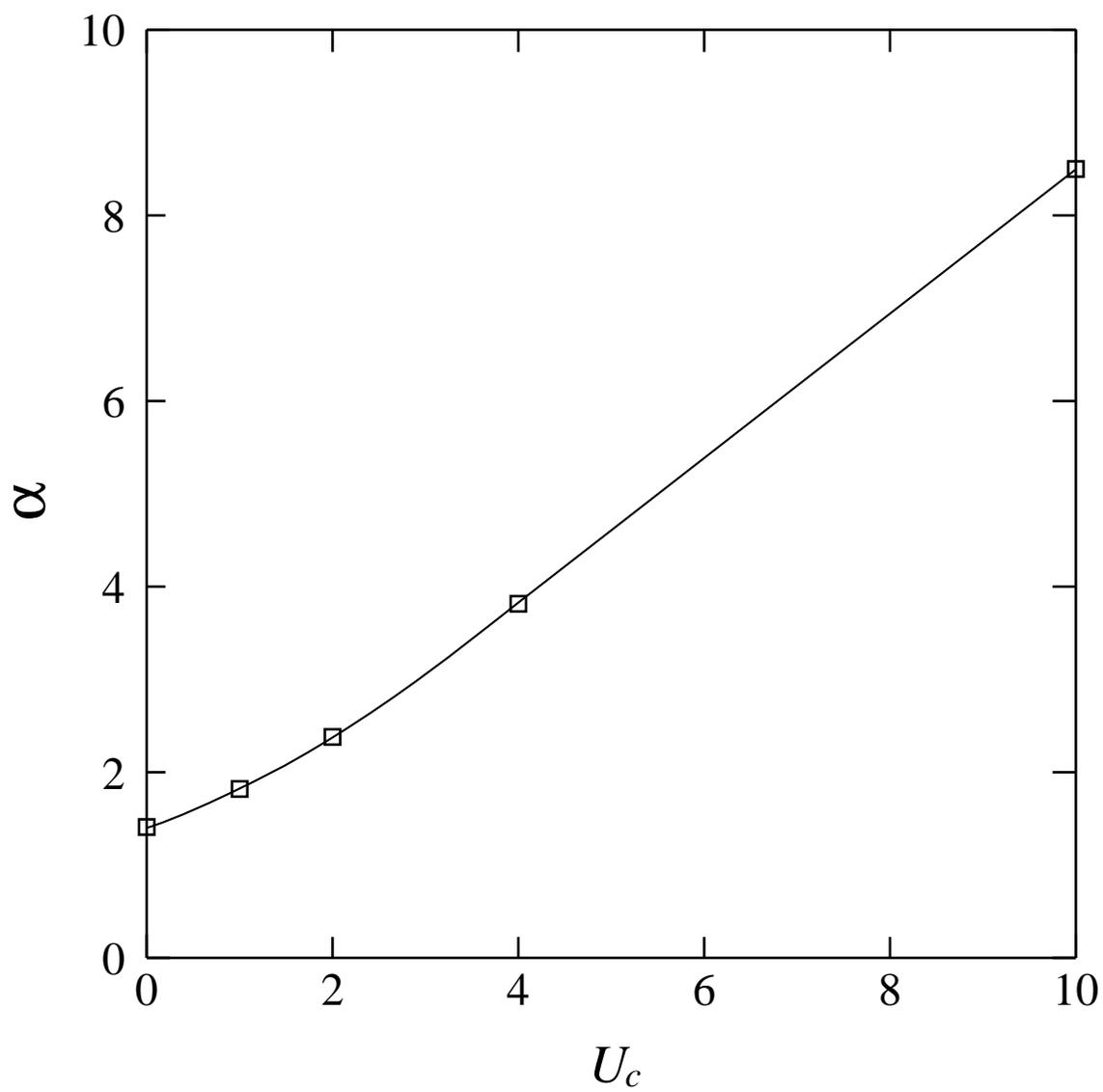
\begin{figure}
\caption{$U_c$-dependence of the exponent of the spin gap.
Coulomb interaction $U_c$ is in units of $t$.}
\label{alpha}
\end{figure}

Now we proceed to the effect of the Coulomb interaction
between conduction electrons.
As is well known, one of the most important effect of
the Coulomb interaction for low-energy properties is the mass 
enhancement, which may be represented by an effective density
of states at the Fermi energy, $\rho^*$. 
Thus in the weak-coupling region it is natural to extend the 
form of Eq.\ (\ref{s_gap}) to finite $U_c$ by allowing 
$U_c$ dependence of the exponent, $\alpha (U_c)$.
In Fig.\ \ref{spin_gap} it is seen that the numerical data
are nicely fitted by this form.
The exponent $\alpha$ is determined for various $U_c$.
As shown in Fig.\ \ref{alpha} the exponent $\alpha$ increases 
with increasing $U_c$ and the asymptotic behavior is 
linear in $U_c$.

In the limit of strong Coulomb interaction 
($U_c/t \rightarrow \infty$) the KLMC is mapped to the Heisenberg 
chain coupled with localized $f$ spins. 
Since the effective coupling of the Heisenberg chain is given by 
$J_{\mbox{\scriptsize eff}}=4t^2/U_c$, we can compare the present
spin gap with that of the spin system.
From Fig.\ \ref{alpha} we find the asymptotic behavior of 
the exponent is $\alpha = 0.78U_c/t + 0.7$. 
It means that the spin gap of the above spin 
system behaves as 
$\Delta_s \sim \exp{(-2J_{\mbox{\scriptsize eff}}/J)}$.
In order to check this form we have analyzed
the numerical data for the spin system obtained 
by Igarashi $et\ al.$ \cite{igarashi} 
and found good agreement.
Thus we conclude that the spin gap of the KLMC always vanishes 
exponentially in the limit of weak exchange coupling for any 
Coulomb interaction, $U_c$, and the exponent $\alpha$ increases 
monotonically with increasing $U_c$.

In the strong-coupling region of $J$, we may use the 
perturbation theory with respect to 
$t/J$ also in the presence of finite Coulomb interaction $U_c$.
After straightforward calculation we 
get the following result for the spin gap:
\begin{equation}
\Delta_s= J-\frac{4t^2}{\frac{1}{2}J+U_c}+\frac{2t^2}
{\frac{3}{2}J+U_c}.
\label{p_s_gap}
\end{equation}
The solid curve in Fig.\ \ref{spin_gap} represents 
this result for $U_c=10t$ which shows that our numerical data 
for the spin gap is well reproduced by the perturbation result
down to $J \sim t$.
From Eq.\ (\ref{p_s_gap}) one can see that the derivative 
of the spin gap with respect to $U_c$ is always positive, 
$ \frac{\partial}{\partial U_c}\Delta_s > 0 $, 
for any positive $U_c$ and $J$. 
From these observations in both the strong- and weak-coupling 
regions we can conclude that the spin gap increases
with increasing Coulomb interaction for any 
antiferromagnetic exchange $J$.

The charge gap is obtained by $E_g(S^z=0,N=L+2)-E_g(S^z=0,N=L)$.
Owing to the hidden SU(2) symmetry in the charge space, 
the energy difference is the same as the charge excitation gap
in the subspace of total electron number fixed to $L$ 
\cite{nishino}.

Before discussing the charge gap 
we first notice the relation between the 
charge gap $\Delta_c$ and the quasiparticle gap $\Delta_{qp}$
which is defined by $E_g(S^z=\pm 1/2,N=L\pm 1)-E_g(S^z=0,N=L)$.
In the strong-coupling limit, $J/t \rightarrow \infty$, 
it is evident that the charge gap is twice
the quasiparticle gap because the energy required to create 
the lowest charge excited state is the same as the
energy cost to add two additional electrons owing to the 
SU(2) symmetry in the charge space. 
In the second order perturbation in $t/J$, one can show 
that the interaction between the two additional electrons is 
repulsive, leading to only a phase shift. 
Therefore, the charge gap is given by the
sum of two quasiparticle gap $\Delta_{qp}$ in the bulk limit:
\begin{equation}
\Delta_c=2\Delta_{qp}.
\end{equation}
A similar argument is also valid for the periodic 
Anderson model \cite{nishino}.
Validity of this relation is checked by the 
present DMRG calculation and we have confirmed this 
relation in the entire region of the exchange constant $J$. 
On the other hand the spin gap is determined by the lowest 
bound state of a quasielectron and a quasihole.

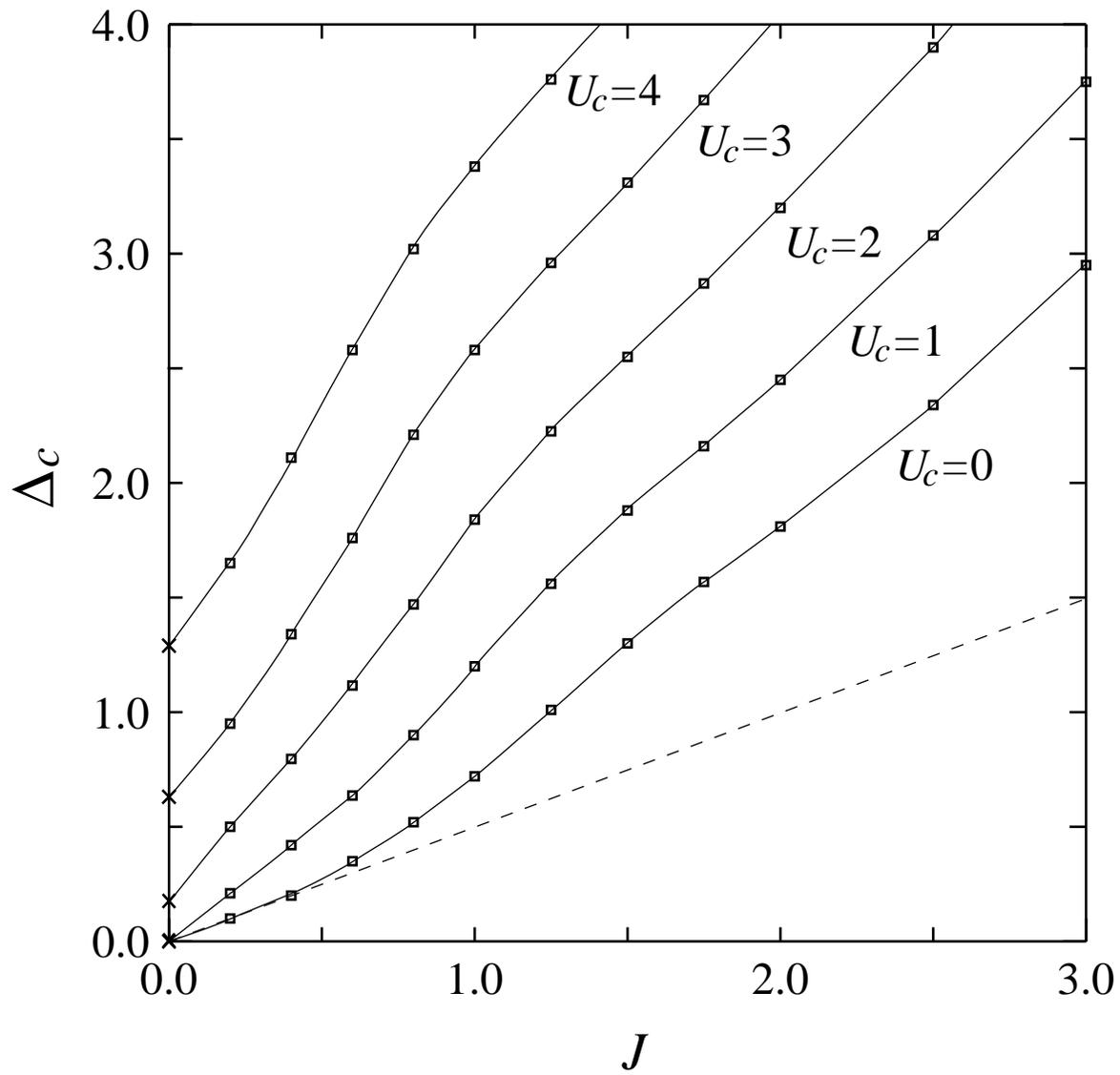
\begin{figure}
\caption{Charge gap of the one-dimensional Kondo lattice model
with Coulomb interaction. Results on the vertical axis are 
obtained from the exact solution of Lieb-Wu.
Typical truncation error in the DMRG calculation is 
$10^{-6}$ for $J=1$ and $10^{-4}$ for $J=0.2$, 
which is a dominant source of numerical errors since
the finite-size scaling, Eq.\ (2), is well obeyed.
Gap energies, exchange constant $J$, 
and Coulomb interaction $U_c$ are in units of $t$.
}
\label{charge_gap}
\end{figure}

Let us start from the case of $J=0$ where exact
results are known. In this limit the KLMC is 
reduced to the Hubbard model which is exactly solved
by Lieb and Wu for the one-dimensional case \cite{lieb}.
The asymptotic form of the charge gap is
$\Delta_c\propto \sqrt{U_ct}\exp
{(-\frac{1}{\rho U_c})}$ for small $U_c$,
and $\Delta_c\propto U_c-4t$ for large $U_c$.
Figure \ref{charge_gap} shows the charge gap extrapolated to 
the infinite system from the data for $L=6,8,12,18,24,40$.
These results for finite $J$
are consistent with the exact results which are 
denoted by crosses on the vertical axis.

For $U_c=0$, the charge gap of the Hubbard model 
vanishes. In this case an important question is how 
the charge gap opens for finite $J$. 
As is shown in the previous work by Nishino $et\ al.$
the charge gap is much bigger than the 
spin gap in the weak-coupling regime \cite{nishino}.
It implies that the correlation length for the spin degrees of 
freedom is much longer than the charge correlation length. 
Therefore, for the discussion of the charge gap it is 
justified to assume that the spin-spin correlation 
length is infinitely long.
Under the assumption of the infinite spin correlation length,
one can get the charge gap which is linear in $J$ with its 
coefficient $1/2$
\begin{equation}
 \Delta_c = \frac{J}{2}.
\end{equation}
This linear dependence in $J$ is actually seen in the
present DMRG calculations shown in Fig.\ \ref{charge_gap}.
For the case of $J=0.2t$ and $U_c=0$ we have additionally 
calculated a larger system of $L=80$
and we get the charge gap of $0.1t$ from the extrapolation.
We may take this fact as a numerical confirmation of the 
coefficient of $1/2$.
It should be stressed again that this type of mean-field 
theory is not justified for the discussion of the spin gap.

From Fig.\ \ref{charge_gap} we also find that 
the charge gap increases with increasing  
Coulomb interaction $U_c$.
In the strong-coupling regime, 
we get the following result for the charge excitation gap
within the second order perturbation:
\begin{equation}
\Delta_c=\frac{3}{2}J+U_c-2t
+\frac{5t^2}{\frac{3}{2}J+U_c}-\frac{3t^2}{J}.
\end{equation}
From the result it is clear that 
the charge gap is a increasing function of $U_c$ for large $J$.
Thus it is concluded that the charge gap increases 
with increasing Coulomb interaction $U_c$ for any exchange
constant $J$ similarly as the spin gap.

The authors would like to thank T.\ Tonegawa for providing
us the data for the Heisenberg chain coupled with localized 
spins \cite{igarashi}.
This work is financially supported by a Grant-in-Aid from 
the Ministry of Education, Science and Culture, Japan.
A part of the numerical calculation was done by VPP500
at the Supercomputer Center of the ISSP, Univ. of Tokyo.
N.\ S.\ is supported by the Japan Society for the 
Promotion of Science.

\end{document}